\documentclass[journal=jacsat,manuscript=article]{achemso}

\usepackage{chemformula} 
\usepackage[T1]{fontenc} 
\usepackage{graphicx}
\usepackage{amsmath}
\usepackage{amssymb}
\usepackage{units}
\usepackage{color}

\DeclareMathOperator{\erf}{erf}

\author{S. Aeschlimann}
\email{sven.aeschlimann@mpsd.mpg.de}
\affiliation{Max Planck Institute for the Structure and Dynamics of Matter, Center for Free Electron Laser Science, Hamburg, Germany}

\author{A. Rossi}
\affiliation{Center for Nanotechnology Innovation at NEST, Italian Institute of Technology, Pisa, Italy}
\alsoaffiliation{NEST, Istituto Nanoscienze, CNR and Scuola Normale Superiore, Pisa, Italy}

\author{M. Ch{\'a}vez-Cervantes}
\affiliation{Max Planck Institute for the Structure and Dynamics of Matter, Center for Free Electron Laser Science, Hamburg, Germany}

\author{R. Krause}
\affiliation{Max Planck Institute for the Structure and Dynamics of Matter, Center for Free Electron Laser Science, Hamburg, Germany}

\author{B. Arnoldi}
\author{B. Stadtm\"uller}
\author{M. Aeschlimann}
\affiliation{Department of Physics and Research Center OPTIMAS, University of Kaiserslautern, Germany}

\author{S. Forti}
\affiliation{Center for Nanotechnology Innovation at NEST, Italian Institute of Technology, Pisa, Italy}

\author{F. Fabbri}
\affiliation{Center for Nanotechnology Innovation at NEST, Italian Institute of Technology, Pisa, Italy}
\alsoaffiliation{NEST, Istituto Nanoscienze, CNR and Scuola Normale Superiore, Pisa, Italy}
\alsoaffiliation{Graphene Labs, Istituto Italiano di Tecnologia, Genova, Italy}

\author{C. Coletti}
\affiliation{Center for Nanotechnology Innovation at NEST, Italian Institute of Technology, Pisa, Italy}
\alsoaffiliation{Graphene Labs, Istituto Italiano di Tecnologia, Genova, Italy}

\author{I. Gierz}
\email{isabella.gierz@mpsd.mpg.de}
\affiliation{Max Planck Institute for the Structure and Dynamics of Matter, Center for Free Electron Laser Science, Hamburg, Germany}

\title{Direct evidence for efficient ultrafast charge separation in epitaxial WS$_2$/graphene heterostructure}

\abbreviations{tr-ARPES}

\begin{document}






\pagebreak

\begin{abstract}
We use time- and angle-resolved photoemission spectroscopy (tr-ARPES) to investigate ultrafast charge transfer in an epitaxial heterostructure made of monolayer WS$_2$ and graphene. This heterostructure combines the benefits of a direct gap semiconductor with strong spin-orbit coupling and strong light-matter interaction with those of a semimetal hosting massless carriers with extremely high mobility and long spin lifetimes. We find that, after photoexcitation at resonance to the A-exciton in WS$_2$, the photoexcited holes rapidly transfer into the graphene layer while the photoexcited electrons remain in the WS$_2$ layer. The resulting charge transfer state is found to have a lifetime of $\sim1$\,ps. We attribute our findings to differences in scattering phase space caused by the relative alignment of WS$_2$ and graphene bands as revealed by high resolution ARPES. In combination with spin-selective excitation using circularly polarized light the investigated WS$_2$/graphene heterostructure might provide a new platform for efficient optical spin injection into graphene.
\end{abstract}

\pagebreak


The availability of many different two-dimensional materials has opened up the possibility to create novel ultimately thin heterostructures with completely new functionalities based on tailored dielectric screening and various proximity-induced effects \cite{GeimNature2013, NovoselovScience2016, LiuNatRevMater2016}. Proof-of-principle devices for future applications in the field of electronics and optoelectronics have been realized \cite{GergiouNatNanotechnol2013, KoppensNatNanotechnol2014, LiPhotonicsRes2017}.

Here we focus on epitaxial van-der-Waals heterostructures consisting of monolayer WS$_2$, a direct-gap semiconductor with strong spin-orbit coupling and a sizable spin splitting of the band structure due to broken inversion symmetry \cite{ZhuPhysRevB2011, WangNatNanotechnol2012}, and monolayer graphene, a semimetal with conical band structure and extremely high carrier mobility \cite{NovoselovScience2004, NovoselovNature2005, CastroNetoRevModPhys2009}, grown on hydrogen-terminated SiC(0001). First indications for ultrafast charge transfer \cite{HeNatCommun2014, HuoJMaterChemC2015, MassicotteNatNanotechnol2016, HillPhysRevB2017, YuanSciAdv2018} and proximity-induced spin-orbit coupling effects \cite{AvsarNatCommun2014, KaloniApplPhysLett2014, WakamuraPhysRevLett2018} make WS$_2$/graphene and similar heterostructures promising candidates for future optoelectronic \cite{RossiNanoscale2018} and optospintronic \cite{OmarPhysRevB2017} applications.

We set out to reveal the relaxation pathways of photogenerated electron-hole pairs in WS$_2$/graphene with time- and angle-resolved photoemission spectroscopy (tr-ARPES). For that purpose we excite the heterostructure with 2\,eV pump pulses resonant to the A-exciton in WS$_2$ \cite{HillPhysRevB2017, SupMat} and eject photoelectrons with a second time-delayed probe pulse at 26\,eV photon energy. We determine kinetic energy and emission angle of the photoelectrons with a hemispherical analyzer as a function of pump-probe delay to get access to the momentum-, energy-, and time-resolved carrier dynamics. The energy and time resolution is 240\,meV and 200\,fs, respectively.

Our results provide direct evidence for ultrafast charge transfer between the epitaxially aligned layers, confirming first indications based on all-optical techniques in similar manually assembled heterostructures with arbitrary azimuthal alignment of the layers \cite{HeNatCommun2014, HuoJMaterChemC2015, MassicotteNatNanotechnol2016, HillPhysRevB2017, YuanSciAdv2018}. In good agreement with Refs. \cite{HeOptExpr2017, SongOptik2018}, we show that this charge transfer is highly asymmetric. Our measurements reveal a charge-separated transient state with photoexcited electrons and holes located in the WS$_2$ and graphene layer, respectively, that lives for $\sim1$\,ps. We interpret our findings in terms of differences in scattering phase space for electron and hole transfer caused by the relative alignment of WS$_2$ and graphene bands as revealed by high resolution ARPES. Combined with spin- and valley-selective optical excitation \cite{YaoPhysRevB2008, XiaoPhysRevLett2012, CaoNatCommun2012, MakNatNanotechnol2012, ZengNanoTechnol2012, XiePNAS2016, BertoniPhysRevLett2016, ChenACSNano2017} WS$_2$/graphene heterostructures might provide a new platform for efficient ultrafast optical spin-injection into graphene.

Details about sample growth and characterization as well as the tr-ARPES setup are provided in the Supporting Information \cite{SupMat}.

\begin{figure}
	\center
		\includegraphics[width = 1\columnwidth]{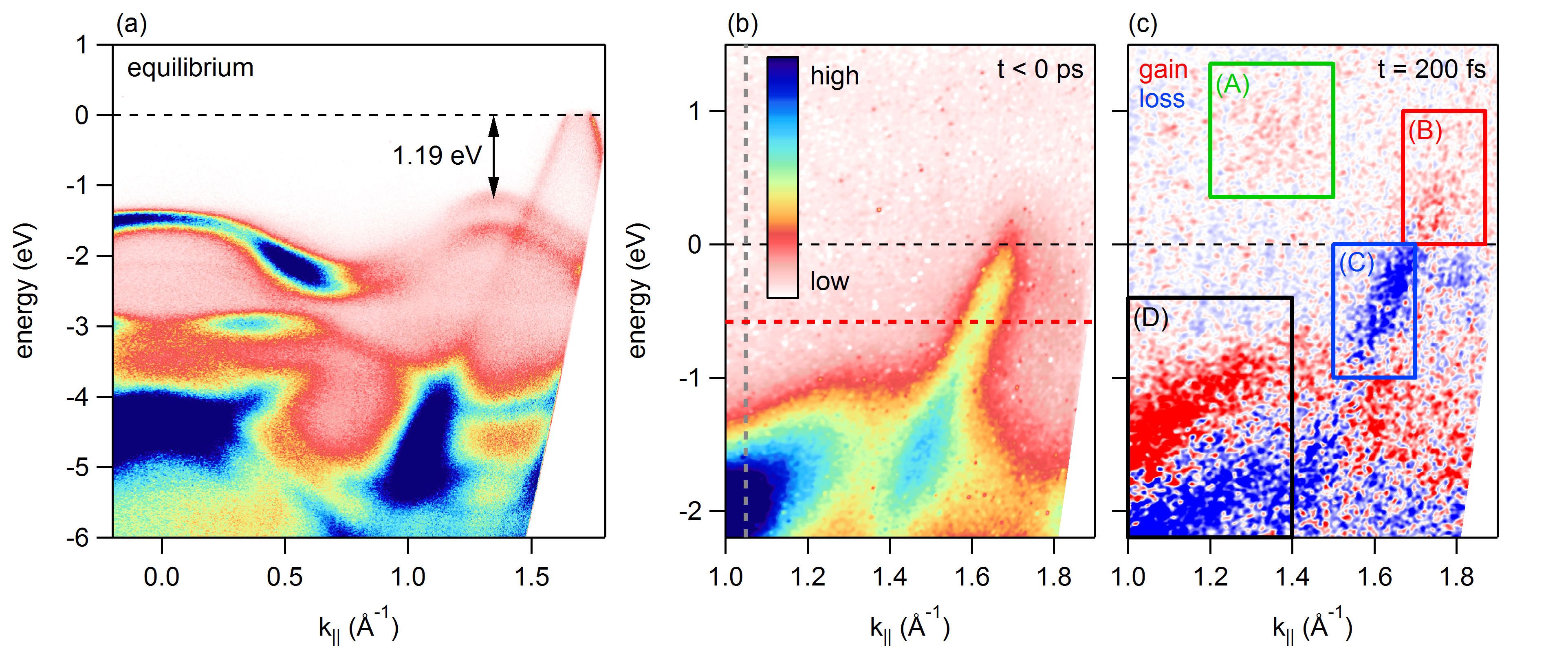}
  \caption{\textbf{Equilibrium band structure and photo-carrier dynamics of WS$_2$/graphene heterostructure.} (a) Equilibrium photocurrent measured along the $\Gamma$K-direction with an unpolarized helium lamp. (b) Photocurrent for negative pump-probe delay measured with p-polarized XUV pulses at 26\,eV photon energy. Dashed gray and red lines mark the position of the line profiles used to extract the transient peak positions in Fig. \ref{fig2}. (c) Pump-induced changes of the photocurrent 200\,fs after photoexcitation at a pump photon energy of 2\,eV with a pump fluence of 2\,mJ/cm$^2$. Gain and loss of photoelectrons are shown in red and blue, respectively. The boxes indicate the area of integration for the pump-probe traces displayed in Fig. \ref{fig3}.}
  \label{fig1}
\end{figure}

Figure \ref{fig1}a shows a high resolution ARPES measurement obtained with a helium lamp of the band structure along the $\Gamma$K-direction of the epitaxial WS$_2$/graphene heterostructure. The Dirac cone is found to be hole-doped with the Dirac point located  $\sim0.3$\,eV above the equilibrium chemical potential. The top of the spin-split WS$_2$ valence band is found to be $\sim1.2$\,eV below the equilibrium chemical potential.

Figure \ref{fig1}b shows a tr-ARPES snapshot of the band structure close to the WS$_2$ and graphene K-points measured at negative pump-probe delay before the arrival of the pump pulse with 100\,fs XUV pulses at 26\,eV photon energy. Here, the spin splitting is not resolved due to the presence of the 2\,eV pump pulse that causes space charge broadening of the spectral features. Figure \ref{fig1}c shows the pump-induced changes of the photocurrent with respect to Fig. \ref{fig1}b at a pump-probe delay of 200\,fs where the pump-probe signal reaches its maximum. Red and blue colors indicate gain and loss of photoelectrons, respectively. 

\begin{figure}
	\center
		\includegraphics[width = 0.7\columnwidth]{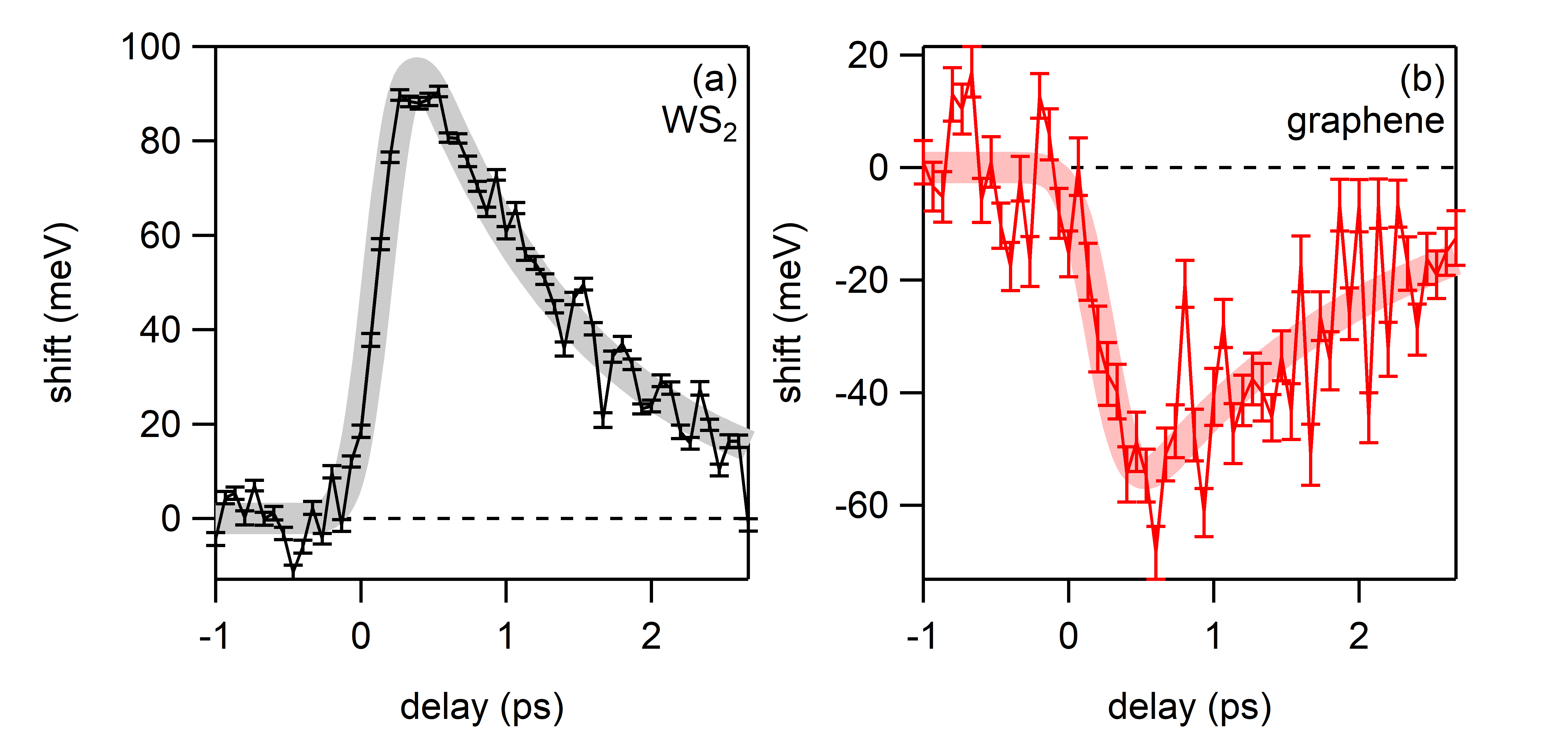}
  \caption{\textbf{Transient band shifts after photoexcitation.} Change in peak position of the WS$_2$ valence band (a) and graphene $\pi$-band (b) as a function of pump-probe delay together with exponential fits. The lifetime of the WS$_2$ shift in (a) is $1.2\pm0.1$\,ps. The lifetime of the graphene shift in (b) is $1.7\pm0.3$\,ps.}
  \label{fig2}
\end{figure}

In order to analyze this rich dynamics in more detail we first determine the transient peak positions of the WS$_2$ valence band and the graphene $\pi$-band along the dashed lines in Fig. \ref{fig1}b as explained in detail in the Supporting Information \cite{SupMat}. We find that the WS$_2$ valence band shifts up by 90\,meV (Fig. \ref{fig2}a) and the graphene $\pi$-band shifts down by 50\,meV (Fig. \ref{fig2}b). The exponential lifetime of these shifts is found to be $1.2\pm0.1$\,ps for the valence band of WS$_2$ and $1.7\pm0.3$\,ps for the graphene $\pi$-band. These peak shifts provide first evidence of a transient charging of the two layers, where additional positive (negative) charge increases (decreases) the binding energy of the electronic states. Note that the up-shift of the WS$_2$ valence band is responsible for the prominent pump-probe signal in the area marked by the black box in Fig. \ref{fig1}c.

\begin{figure}
	\center
		\includegraphics[width = 1\columnwidth]{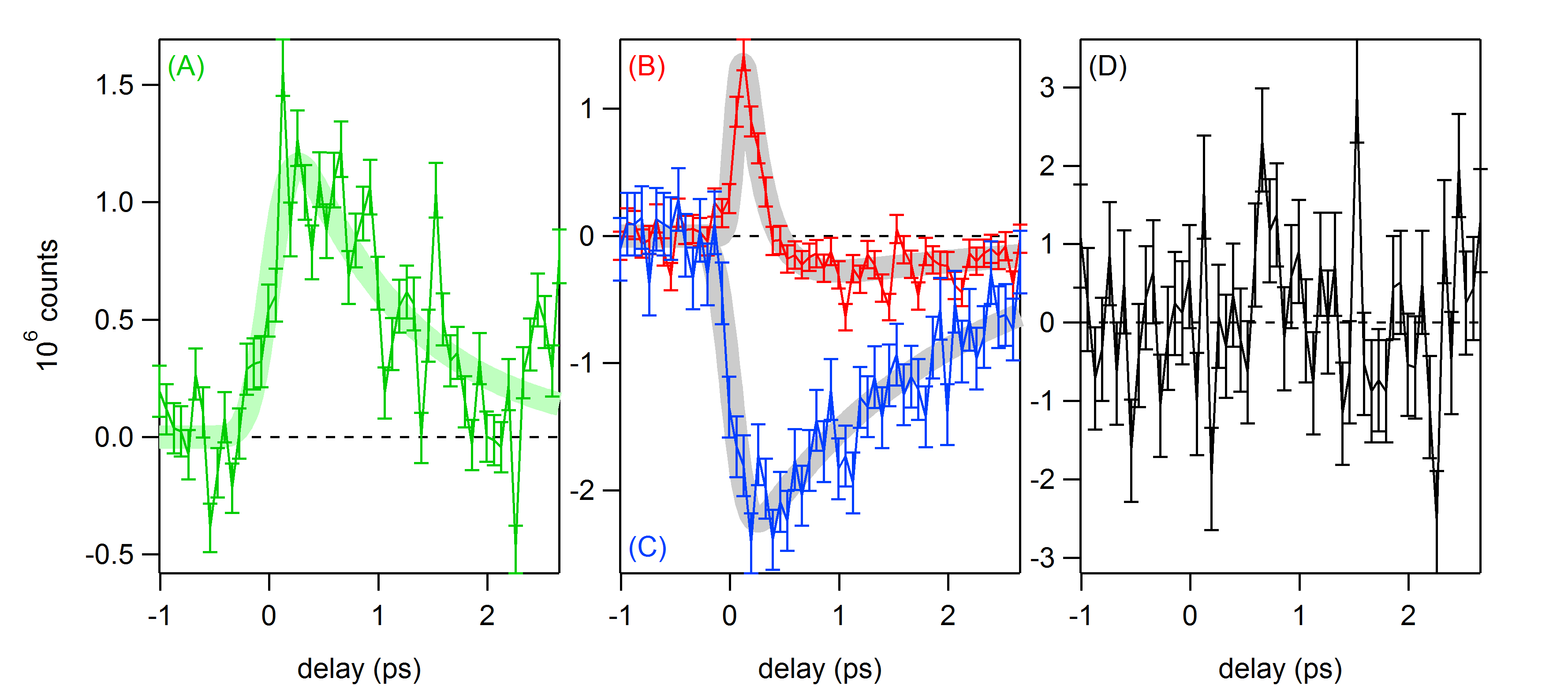}
  \caption{\textbf{Energy- and momentum-resolved carrier dynamics.} Pump-probe traces as a function of delay obtained by integrating the photocurrent over the area indicated by the boxes in Fig. \ref{fig1}c. The thick lines are exponential fits to the data. (A) Transient carrier population in the conduction band of WS$_2$. (B) Pump-probe signal of the $\pi$-band of graphene above the equilibrium chemical potential. (C) Pump-probe signal of the $\pi$-band of graphene below the equilibrium chemical potential. (D) Net pump-probe signal in the valence band of WS$_2$. The lifetimes are found to be $1.2\pm0.1$\,ps in (A), $180\pm20$\,fs (gain) and $\sim2$\,ps (loss) in (B), and $1.8\pm0.2$\,ps in (C).}
  \label{fig3}
\end{figure}

Next we integrate the pump-probe signal over the areas indicated by the colored boxes in Fig. \ref{fig1}c and plot the resulting counts as a function of pump-probe delay in Fig. \ref{fig3}. Figure \ref{fig3}A shows the dynamics of the photoexcited carriers close to the bottom of the conduction band of the WS$_2$ layer with a lifetime of $1.1\pm0.1$\,ps obtained from an exponential fit to the data (see Supporting Information \cite{SupMat}). 

In Figs. \ref{fig3}B and C we show the pump-probe signal of the graphene $\pi$-band. Interestingly, we find that the gain of electrons above the equilibrium chemical potential (Fig. \ref{fig3}B) has a much shorter lifetime ($180\pm20$\,fs) compared to the loss of electrons below the equilibrium chemical potential ($1.8\pm0.2$\,ps in Fig. \ref{fig3}C). Further, the initial gain of the photocurrent in Fig. \ref{fig3}B is found to turn into loss at $t=400$\,fs with a lifetime of $\sim2$\,ps. The asymmetry between gain and loss is found to be absent in the pump-probe signal of uncovered monolayer graphene (see Fig. 10 in the Supporting Information \cite{SupMat}), indicating that the asymmetry is a consequence of interlayer coupling in the WS$_2$/graphene heterostructure. The observation of a short-lived gain and long-lived loss above and below the equilibrium chemical potential, respectively, indicates that electrons are efficiently removed from the graphene layer upon photoexcitation of the heterostructure. As a result the graphene layer becomes positively charged which is consistent with the increase in binding energy of the $\pi$-band found in Fig. \ref{fig2}b. The down-shift of the $\pi$-band removes the high-energy tail of the equilibrium Fermi-Dirac distribution from above the equilibrium chemical potential which partly explains the change of sign of the pump-probe signal in Fig. \ref{fig3}B. We will show below that this effect is further enhanced by the transient loss of electrons in the $\pi$-band.

This scenario is supported by the net pump-probe signal of the WS$_2$ valence band in Fig. \ref{fig3}D. This data was obtained by integrating the counts over the area given by the black box in Fig. \ref{fig1}b that captures the electrons photo-emitted from the valence band at all pump-probe delays. Within the experimental error bars we find no indication for the presence of holes in the valence band of WS$_2$ for any pump-probe delay. This indicates that, after photoexcitation, these holes are rapidly refilled on a time scale short compared to our temporal resolution.

\begin{figure}
	\center
		\includegraphics[width = 0.5\columnwidth]{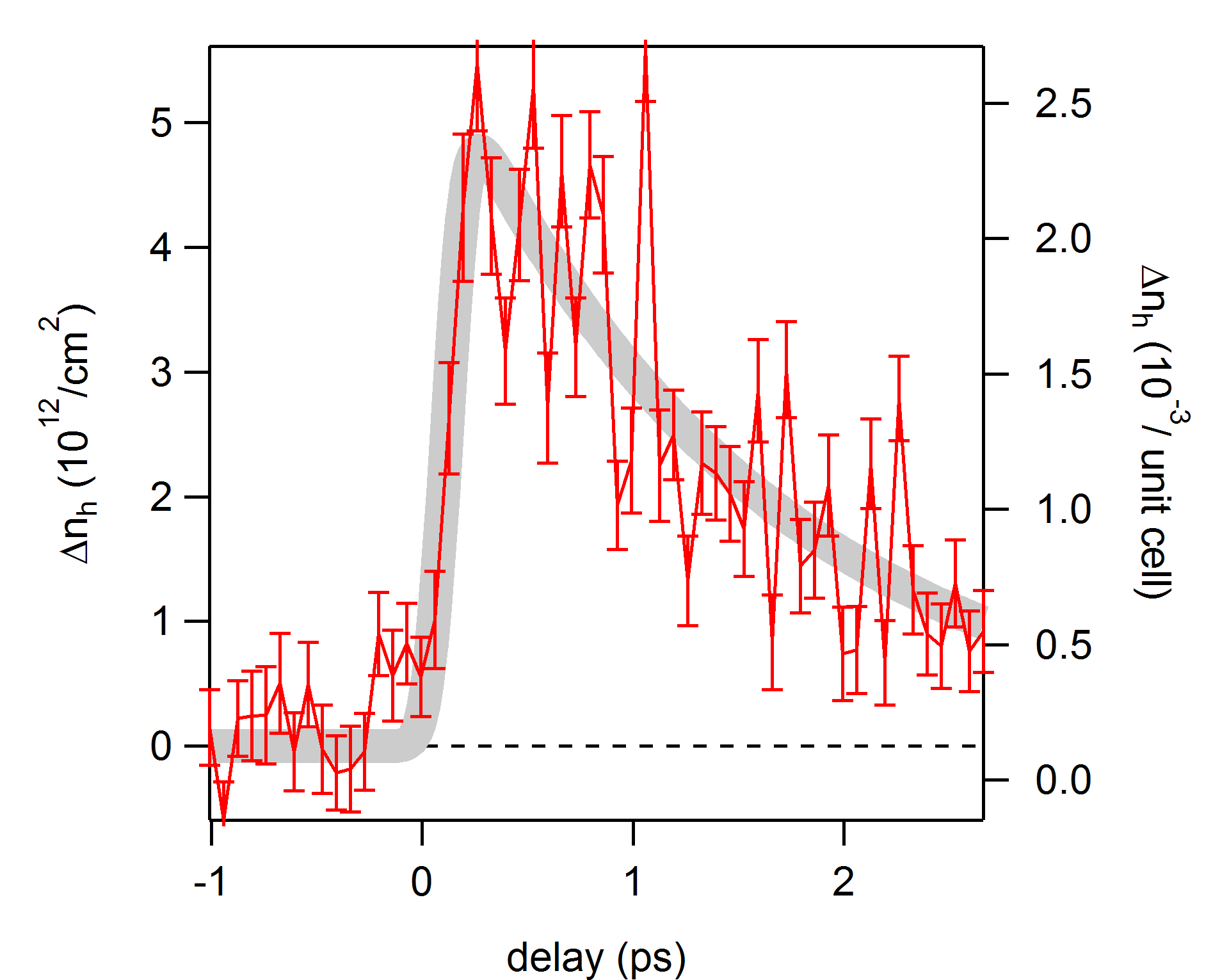}
  \caption{\textbf{Transient hole density in graphene layer.} Change of the number of holes in the $\pi$-band as a function of pump-probe delay together with exponential fit yielding a lifetime of $1.5\pm0.2$\,ps.}
  \label{fig4}
\end{figure}

In order to provide final proof for our hypothesis of ultrafast charge separation in the WS$_2$/graphene heterostructure we determine the number of holes transferred to the graphene layer as described in detail in the Supporting Information \cite{SupMat}. In short, the transient electronic distribution of the $\pi$-band was fitted with a Fermi-Dirac distribution. The number of holes was then calculated from the resulting values for the transient chemical potential and electronic temperature. The result is shown in Fig. \ref{fig4}. We find that a total number of $\sim5\times10^{12}$\,holes/cm$^2$ are transferred from WS$_2$ to graphene with an exponential lifetime of $1.5\pm0.2$\,ps.

\begin{figure}
	\center
		\includegraphics[width = 0.7\columnwidth]{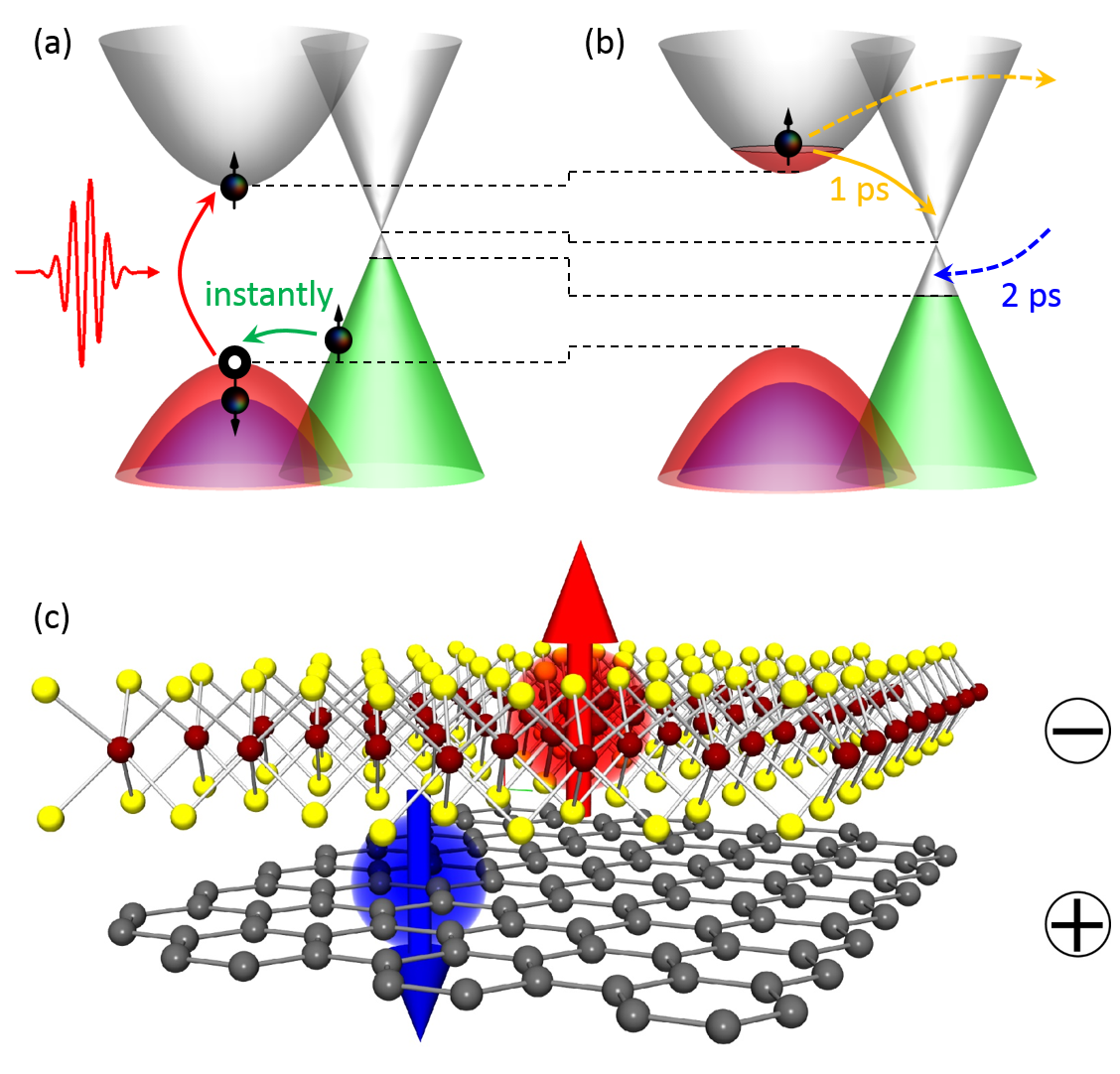}
  \caption{\textbf{Sketch of ultrafast charge transfer deduced from tr-ARPES data.} (a) Photo-excitation at resonance to the WS$_2$ A-exciton at 2\,eV injects electrons into the conduction band of WS$_2$. The corresponding holes in the valence band of WS$_2$ are instantly refilled by electrons from the graphene $\pi$-band. (b) The photoexcited carriers in the conduction band of WS$_2$ have a lifetime of $\sim1$\,ps. The holes in the graphene $\pi$-band live for $\sim2$\,ps, indicating the importance of additional scattering channels indicated by dashed arrows. Black dashed lines in (a) and (b) indicate band shifts and changes in chemical potential. (c) In the transient state the WS$_2$ layer is negatively charged while the graphene layer is positively charged. For spin-selective excitation with circularly polarized light the photoexcited electrons in WS$_2$ and the corresponding holes in graphene are expected to show opposite spin polarization.}
  \label{fig5}
\end{figure}

From the findings in Figs. \ref{fig2}-\ref{fig4} the following microscopic picture for the ultrafast charge transfer in the WS$_2$/graphene heterostructure emerges (Fig. \ref{fig5}). Photo-excitation of the WS$_2$/graphene heterostructure at 2\,eV dominantly populates the A-exciton in WS$_2$ (Fig. \ref{fig5}a). Additional electronic excitations across the Dirac point in graphene as well as between WS$_2$ and graphene bands are energetically possible but considerably less efficient. The photoexcited holes in the valence band of WS$_2$ are refilled by electrons originating from the graphene $\pi$-band on a timescale short compared to our temporal resolution (Fig. \ref{fig5}a). The photoexcited electrons in the conduction band of WS$_2$ have a lifetime of $\sim1$\,ps (Fig. \ref{fig5}b). However, it takes $\sim2$\,ps to refill the holes in the graphene $\pi$-band (Fig. \ref{fig5}b). This indicates that, aside from direct electron transfer between the WS$_2$ conduction band and the graphene $\pi$-band, additional relaxation pathways --- possibly via defect states \cite{CarozoSciAdv2017} --- need to be considered to understand the full dynamics.

In the transient state the photoexcited electrons reside in the conduction band of WS$_2$ while the photoexcited holes are located in the $\pi$-band of graphene (Fig. \ref{fig5}c). This means that the WS$_2$ layer is negatively charged and the graphene layer is positively charged. This accounts for the transient peak shifts (Fig. \ref{fig2}), the asymmetry of the graphene pump-probe signal (Figs. \ref{fig3}B and C), the absence of holes in the valence band of WS$_2$ (Fig. \ref{fig3}D), as well as the additional holes in the graphene $\pi$-band (Fig. \ref{fig4}). The lifetime of this charge-separated state is $\sim1$\,ps (Fig. \ref{fig3}A).

Similar charge-separated transient states have been observed in related van-der-Waals heterostructures made out of two direct-gap semiconductors with type II band alignment and corresponding staggered band gap \cite{HongNatNanotechnol2014, CeballosACSNano2014, RiveraNatCommun2015, ZhuJAmChemSoc2015, LinNatCommun2015, ChenNatCommun2016}. After photoexcitation the electrons and holes were found to rapidly move to the bottom of the conduction band and to the top of the valence band, respectively, that are located in different layers of the heterostructure \cite{HongNatNanotechnol2014, CeballosACSNano2014, RiveraNatCommun2015, ZhuJAmChemSoc2015, LinNatCommun2015, ChenNatCommun2016}. 

In the case of our WS$_2$/graphene heterostructure the energetically most favorable location for both electrons and holes is at the Fermi level in the metallic graphene layer. Therefore, one would expect that both electrons and holes rapidly transfer to the graphene $\pi$-band. However, our measurements clearly show that hole transfer ($<200$\,fs) is much more efficient than electron transfer ($\sim1$\,ps). We attribute this to the relative energetic alignment of the WS$_2$ and the graphene bands as revealed in Fig. \ref{fig1}a that offers a larger number of available final states for hole transfer compared to electron transfer. In the present case, assuming a $\sim2$\,eV WS$_2$ band gap, the graphene Dirac point and equilibrium chemical potential are located $\sim0.5$\,eV above and $\sim0.2$\,eV above the middle of the WS$_2$ band gap, respectively, breaking electron-hole symmetry. We find that the number of available final states for hole transfer is $\sim6$ times larger than for electron transfer (see Supporting Information \cite{SupMat}) which is why hole transfer is expected to be faster than electron transfer.

A complete microscopic picture of the observed ultrafast asymmetric charge transfer should, however, also consider the overlap between the orbitals that constitute the A-exciton wavefunction in WS$_2$ and the graphene $\pi$-band, respectively, different electron-electron and electron-phonon scattering channels including the constraints imposed by momentum, energy, spin, and pseudospin conservation, the influence of plasma oscillations \cite{WangNatCommun2016}, as well as the role of a possible displacive excitation of coherent phonon oscillations that might mediate the charge transfer \cite{ZhengNanoLett2017, ZhengPhysRevB2018}. Also, one might speculate whether the observed charge transfer state consists of charge transfer excitons or free electron-hole pairs (see Supporting Information \cite{SupMat}). Further theoretical investigations that go beyond the scope of the present paper are required to clarify these issues.

In summary we have used time- and angle-resolved photoemission spectroscopy to study ultrafast interlayer charge transfer in an epitaxial WS$_2$/graphene heterostructure. We found that, when excited at resonance to the A-exciton of WS$_2$ at 2\,eV, the photoexcited holes rapidly transfer into the graphene layer while the photoexcited electrons remain in the WS$_2$ layer. We attributed this to the fact that the number of available final states for hole transfer is larger than for electron transfer. The lifetime of the charge-separated transient state was found to be $\sim1$\,ps. In combination with spin-selective optical excitation using circularly polarized light \cite{XiaoPhysRevLett2012, XiePNAS2016, ChenACSNano2017} the observed ultrafast charge transfer might be accompanied by spin transfer. In this case, the investigated WS$_2$/graphene heterostructure might be used for efficient optical spin injection into graphene resulting in novel optospintronic devices.

\begin{acknowledgement}
This work received financial support from the German Sicence Foundation via the Collaborative Research Centers 925 ``Light induced dynamics and control of correlated quantum systems'' (project B6) and 173 ``Spin + X: spin in its collective environment'' (project A02) and from the European Union's Horizon 2020 research and innovation program under grant agreement Nos. 696656 - GrapheneCore1 and 785219 - GrapheneCore2.
\end{acknowledgement}

\begin{suppinfo}
The Supporting Information contains information about sample growth and characterization, the tr-ARPES setup, tr-ARPES data analysis, and further tr-ARPES data on graphene/H-SiC(0001).
\end{suppinfo}


\clearpage
\pagebreak

\section{Supporting Information}

\subsection{Sample Growth and Characterization}


\noindent The graphene samples were grown on commercial semiconducting 6H-SiC(0001) wafers from SiCrystal GmbH. The N-doped wafers were on-axis with a miscut below 0.5$^{\circ}$.

\noindent The SiC substrate was hydrogen etched to remove scratches and obtain regular flat terraces. The clean and atomically flat Si-terminated surface was then graphitized by annealing the sample in Ar atmosphere at 1300$^{\circ}$C for 8 minutes \cite{EmtsevNatMater2009}. In this way, we obtained a single carbon layer where every third carbon atom formed a covalent bond to the SiC substrate \cite{RiedlPhysRevB2007, EmtsevPhysRevB2008, GolerCarbon2013, FortiJPhysDApplPhys2014}. This layer was then turned into completely sp$^2$-hybridized quasi free-standing hole-doped graphene via hydrogen intercalation \cite{RiedlPhysRevLett2009, FortiPhysRevB2011}. These samples are referred to as graphene/H-SiC(0001) in the following. The whole process was carried out in a commercial Black Magic™ growth chamber from Aixtron. 

\noindent The WS$_2$ growth was carried out in a standard hot-wall reactor by low pressure chemical vapor deposition (LPCVD) \cite{Rossi2DMater2016, FortiNanoscale2017} using WO$_3$ and S powders with a mass ratio of 1:100 as precursors. The WO$_3$ and S powders were kept at 900$^{\circ}$C and 200$^{\circ}$C, respectively. The WO$_3$ powder was placed close to the substrate. Argon was employed as carrier gas with a flow of 8\,sccm. The pressure in the reactor was kept at 0.5\,mbar.

\noindent The samples were characterized with secondary electron microscopy (SEM, ZEISS Merlin), atomic force microscopy (Anasys AFM), Raman, and photoluminescence spectroscopy (Renishaw, InVia), as well as low energy electron diffraction (LEED). 

\begin{figure}
	\center
		\includegraphics[width = 1\columnwidth]{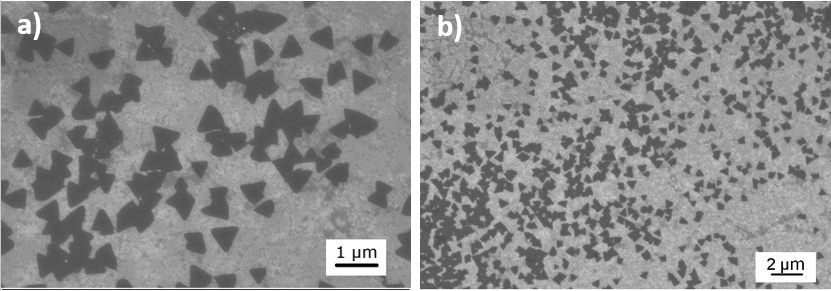}
  \caption{\textbf{SEM analysis.} The pictures were obtained with an accelerating voltage of 5\,keV and a beam current of 30\,pA.}
  \label{figS1}
\end{figure}

\noindent SEM (Fig. \ref{figS1}a) reveals WS$_2$ single-crystalline domains with a side length varying between 300 and 700\,nm. In some areas several single crystalline domains are found to merge.  From the low magnification image in Fig. \ref{figS1}b we estimate a WS$_2$ coverage of 40\% suitable for the tr-ARPES analysis. The orientation of the WS$_2$ triangles reveals the presence of two different domains with an angle of 60$^{\circ}$ between them.

\begin{figure}
	\center
		\includegraphics[width = 0.5\columnwidth]{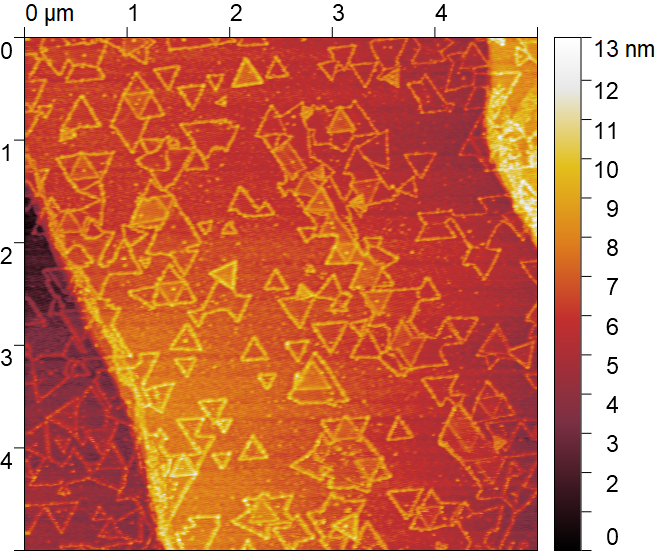}
  \caption{\textbf{AFM analysis.} }
  \label{figS2}
\end{figure}

\noindent The AFM analysis in Fig. \ref{figS2} confirms the island size distribution obtained by SEM. In addition, the topographical map reveals that about 10\% of the flakes consist of WS$_2$ bilayers.

\begin{figure}
	\center
		\includegraphics[width = 1\columnwidth]{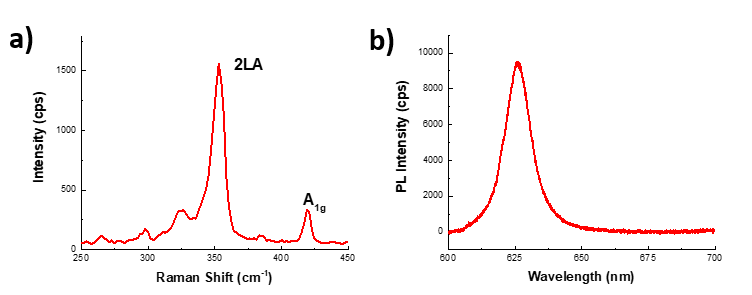}
  \caption{\textbf{Raman (a) and photoluminescence spectrum (b).} The measurements were performed with 1\,mW  532\,nm laser excitation using a 100x-N.A.0.90 objective at room temperature.}
  \label{figS3}
\end{figure}

\noindent The ratio of the 2LA and A1g peak intensities in the Raman spectrum in Fig. \ref{figS3}a confirms the presence of monolayer WS$_2$ \cite{Rossi2DMater2016}. Figure \ref{figS3}b shows the photoluminescence spectrum of the heterostructure with a sharp emission peak at 625\,nm (1.98\,eV) attributed to the WS$_2$ A-exciton \cite{Rossi2DMater2016}.

\begin{figure}
	\center
		\includegraphics[width = 0.3\columnwidth]{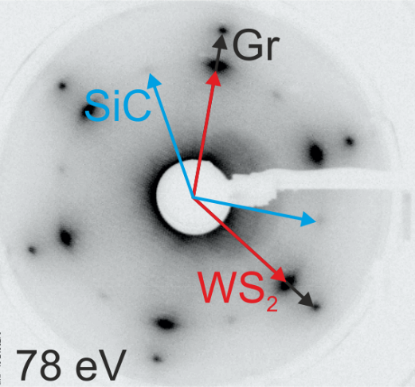}
  \caption{\textbf{LEED.} The picture was obtained with an electron energy of 78\,eV. Blue, black, and red arrows point to the diffraction spots of the SiC substrate, the graphene layer, and the WS$_2$ layer, respectively. }
  \label{figS4}
\end{figure}

\noindent In order to assess the crystalline quality of the samples as well as the relative alignment of the WS$_2$ with respect to the graphene layer LEED measurements were carried out. Prior to the measurement the sample was mildly annealed at 200$^{\circ}$C in ultra-high vacuum. The LEED picture in Fig. \ref{figS4} confirms the perfect azimuthal alignment of the two layers. In combination with the SEM results from Fig. \ref{figS1} we conclude that the WS$_2$ islands grow such that either the $\Gamma$K- or the $\Gamma$K'-direction of the WS$_2$ island is aligned with the $\Gamma$K-direction of the graphene layer.

\subsection*{Tr-ARPES setup}

\noindent The time- and angle-resolved photoemission spectroscopy (tr-ARPES) experiments were performed at the Max Planck Institute for the Structure and Dynamics of Matter in Hamburg. The setup was based on a 1\,kHz Titanium:Sapphire amplifier (Coherent Legend Elite Duo). 2\,mJ of output power were used for high harmonics generation in argon. The resulting extreme ultraviolet light passed through a grating monochromator producing 100\,fs probe pulses at 26\,eV photon energy. 8\,mJ of amplifier output power were sent into an optical parametric amplifier (HE-TOPAS from Light Conversion). The signal beam at 1\,eV photon energy was frequency-doubled in a BBO crystal to obtain the 2\,eV pump pulses. The ARPES measurements were performed with a hemispherical analyzer (Specs Phoibos 100). The overall energy and temporal resolution was 240\,meV and 200\,fs, respectively.

\noindent ARPES is an extremely surface sensitive technique. In order to eject photoelectrons from both the WS$_2$ and the graphene layer, samples with an incomplete WS$_2$ coverage of $\sim40$\% were used.

\subsection*{Photo-carrier dynamics of quasi free-standing graphene}

\begin{figure}
	\center
		\includegraphics[width = 1\columnwidth]{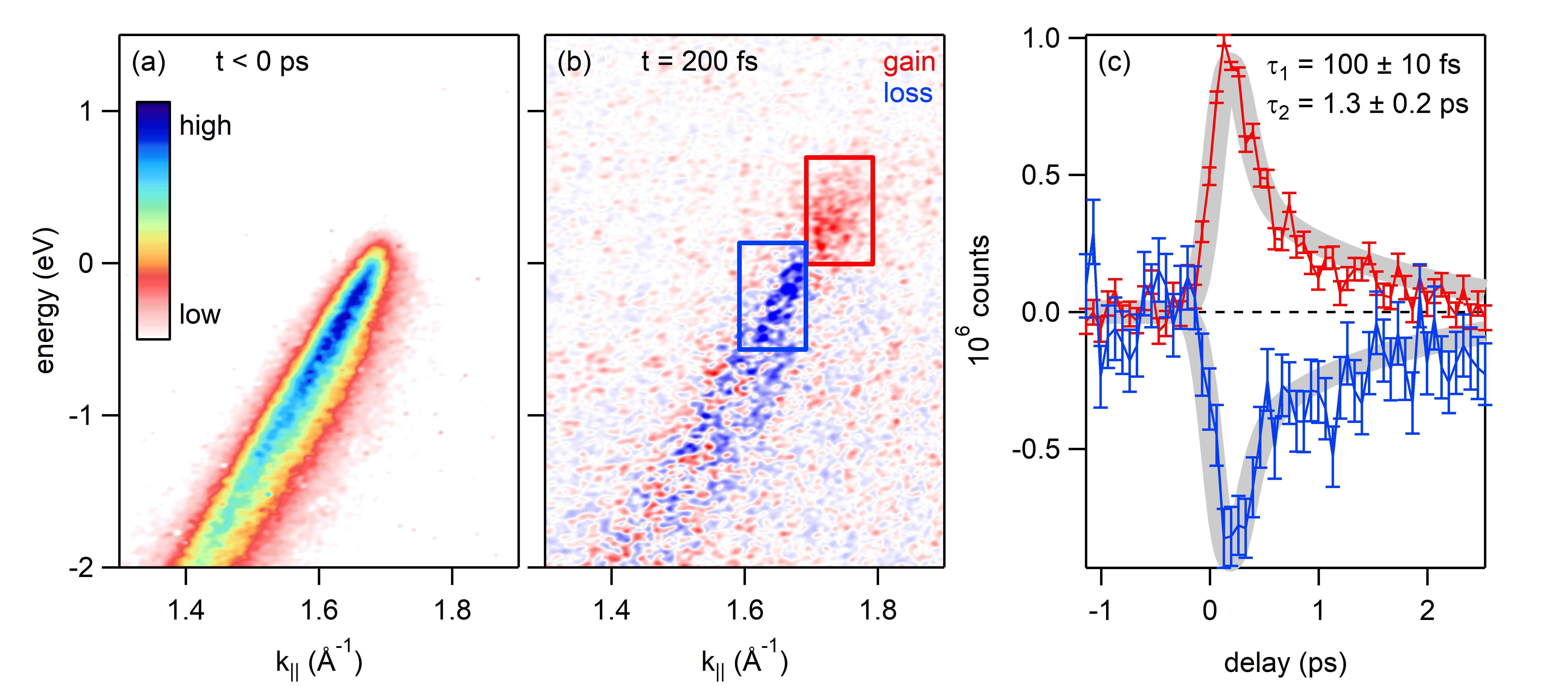}
  \caption{\textbf{Photo-carrier dynamics in graphene/H-SiC(0001).} (a) Photocurrent along the $\Gamma$K-direction for negative pump-probe delay. (b) Pump-induced changes of the photocurrent 200\,fs after photo-excitation at a pump photon energy of 2\,eV with a pump fluence of 2\,mJ/cm$^2$. Gain and loss of photoelectrons are shown in red and blue, respectively. (c) Pump-probe traces as a function of delay obtained by integrating the photocurrent over the area indicated by the red and blue boxes in panel (b). Thick gray lines are double-exponential fits to the data.}
  \label{figS5}
\end{figure}

\noindent In Fig. \ref{figS5} we present tr-ARPES results for graphene/H-SiC(0001) without WS$_2$ on top for the same excitation conditions as the ones for the measurements on the WS$_2$/graphene/H-SiC(0001)  heterostructure in the main text. In contrast to the heterostructure (Fig. 2b and c in the main paper) the gain and loss signal for pure graphene in Fig. \ref{figS5}c is symmetric.

\subsection*{Tr-ARPES data analysis}

\noindent We used the following fitting function to extract rise and decay times from the data presented in Figs. 2-4 of the main text:

\begin{equation*}
f(t)=\frac{a}{2} \left(1+\erf\left(\frac{(t-t_0)\tau-\text{FWHM}^2/(8\ln2)}{\text{FWHM}\,\tau/(2\sqrt{\ln2})}\right)\right) \exp\left(\frac{\text{FWHM}^2/(8\ln2)-2(t-t_0)\tau}{2\tau^2}\right)
\end{equation*}

\noindent $a$ is the amplitude of the pump-probe signal, $\text{FWHM}$ ist the full width at half maximum of the derivative of the rising edge, $t_0$ is the middle of the rising edge, $\erf$ is the error function, and $\tau$ is the exponential lifetime. This fitting function is obtained by convolving the product of a step function and an exponential decay with a Gaussian to account for the finite rise time of the signal. 

\begin{figure}
	\center
		\includegraphics[width = 1\columnwidth]{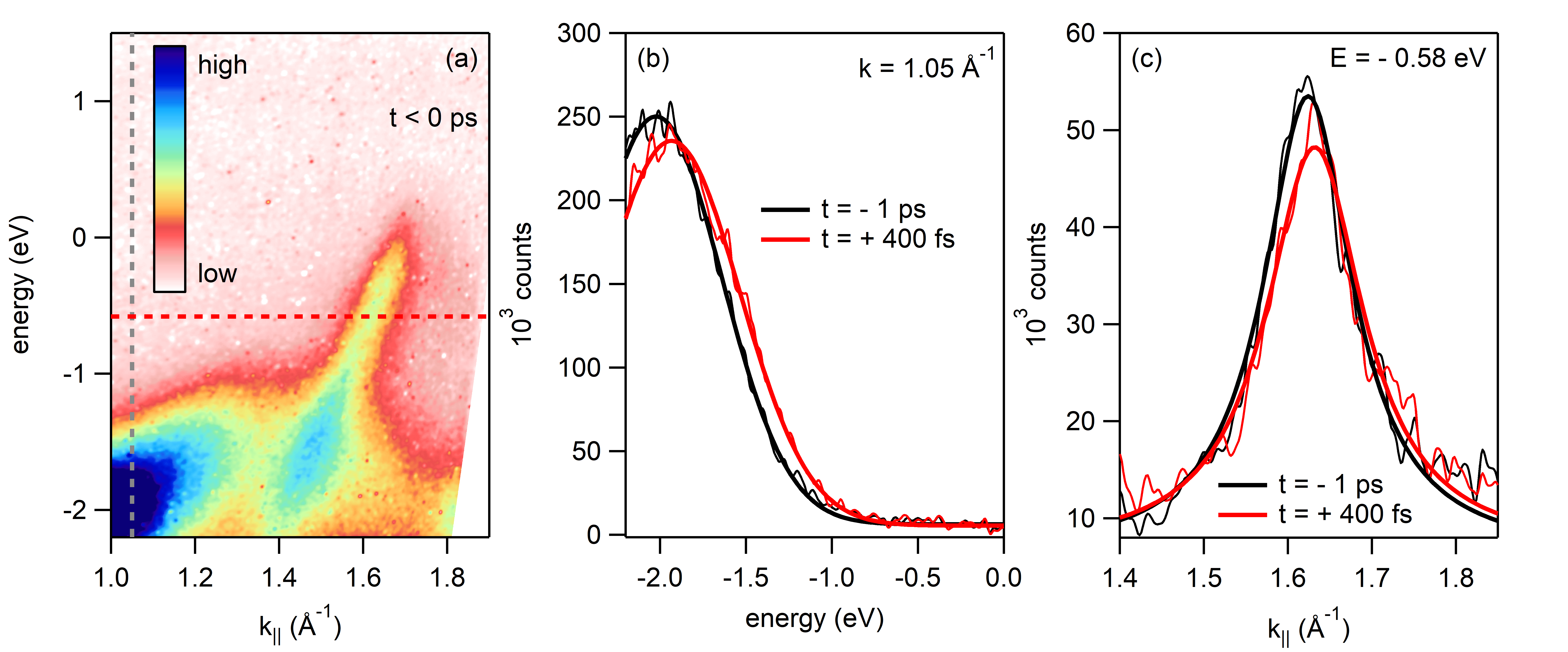}
  \caption{\textbf{Fits of transient peak position.} (a) Photocurrent along the $\Gamma$K-direction for negative pump-probe delay. (b) Energy distribution curves extracted along the dashed grey line in (a) together with Gaussian fits. (c) Momentum distribution curves extracted along the dashed red line in (a) together with Lorentzian fits.}
  \label{figS6}
\end{figure}

\noindent The transient peak shifts in Fig. 3 in the main text were obtained as follows. In order to determine the position of the WS$_2$ valence band we extracted energy distribution curves (EDCs) at $k_{||}=1.05$\AA$^{-1}$ from Fig. \ref{figS6}a and fitted them with Gaussians (Fig. \ref{figS6}b). The transient position of the graphene $\pi$-band was obtained by fitting momentum distribution curves (MDCs) at $E=-0.58$\,eV from Fig. \ref{figS6}a with Lorentzians (Fig. \ref{figS6}c). The MDC position was then multiplied with the slope of the $\pi$-band $\hbar v_F$=7\,eV\AA, yielding the transient band shift in Fig. 3b in the main text.

\begin{figure}
	\center
		\includegraphics[width = 1\columnwidth]{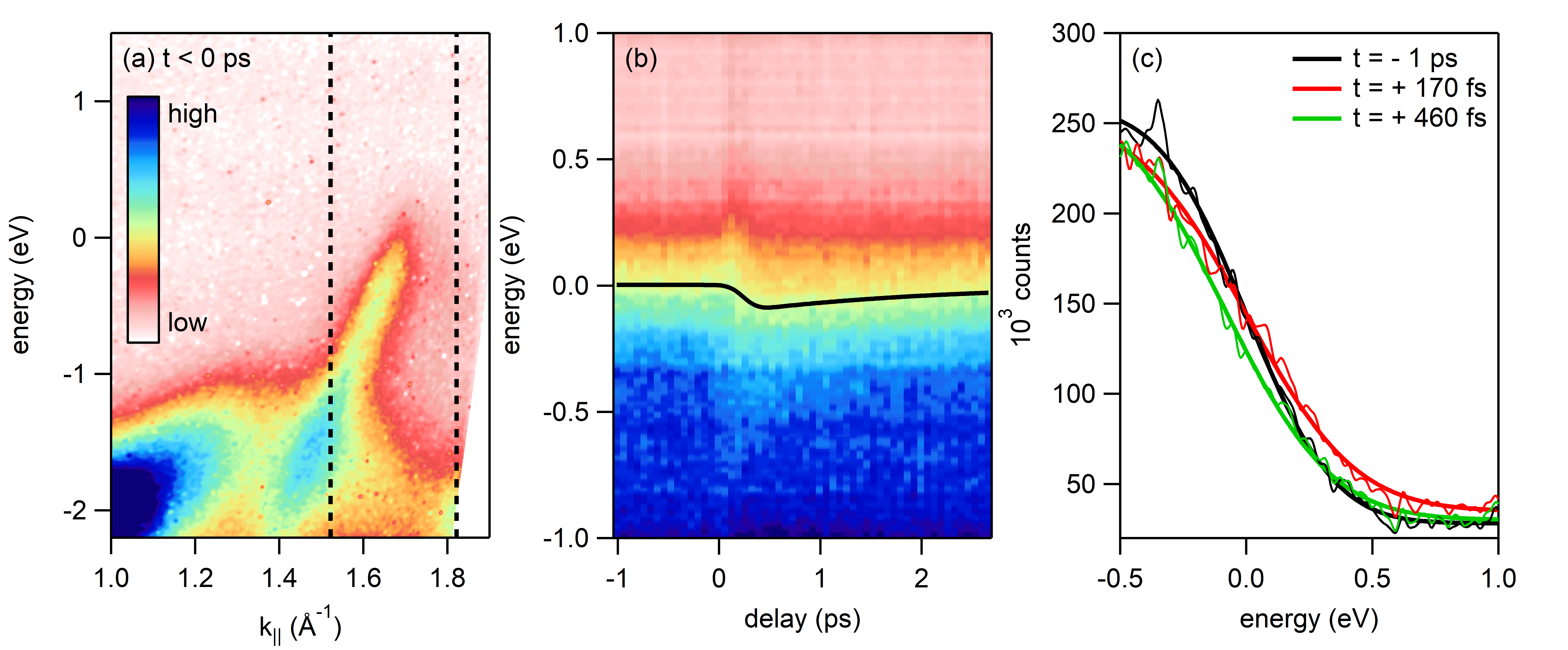}
  \caption{\textbf{Fermi-Dirac fits for graphene $\pi$-band.} (a) Photocurrent along the $\Gamma$K-direction for negative pump-probe delay. (b) Photocurrent integrated over the area between the two dashed lines in (a) as a function of pump-probe delay. The black line indicates the transient position of the chemical potential referenced to the vacuum level $\mu_{e (vac)}$. (c) Momentum-integrated energy distribution curves from (b) for three different time delays together with Fermi-Dirac fits.}
  \label{figS7}
\end{figure}

\begin{figure}
	\center
		\includegraphics[width = 1\columnwidth]{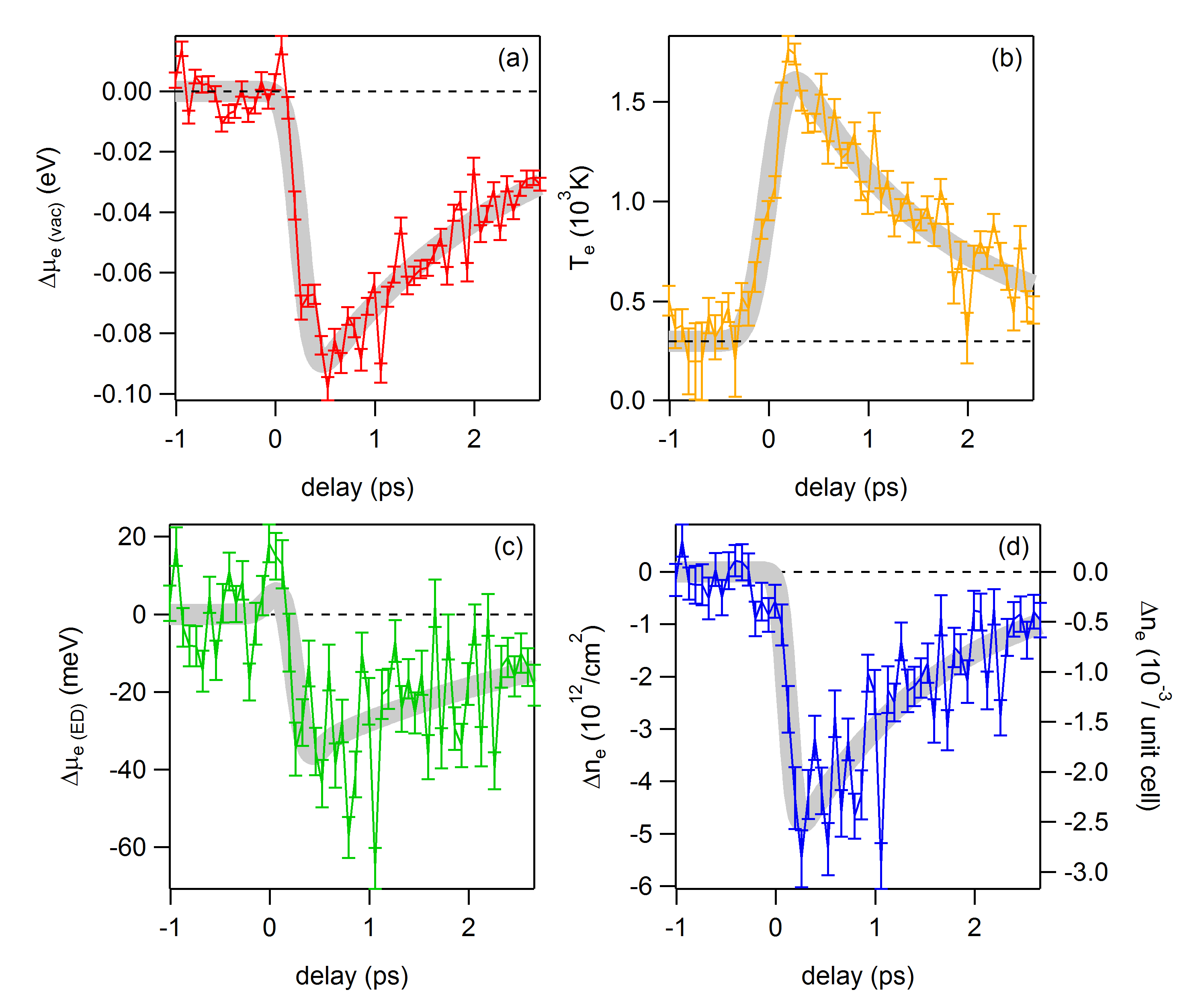}
  \caption{\textbf{Result of Fermi-Dirac fits for graphene $\pi$-band.} (a) Change of the chemical potential referenced to the vacuum level $\mu_{e (vac)}$ as a function of pump-probe delay. (b) Electronic temperature as a function of pump-probe delay. Thick lines in (a) and (b) are exponential fits with lifetimes of $2.0\pm0.2$\,ps and $1.5\pm0.1$\,ps, respectively. (c) Shift of the chemical potential referenced to the Dirac point $\mu_{e (ED)}$ obtained by subtracting the $\pi$-band shift from Fig. 2b from the main text from $\mu_{e (vac)}$ in panel (a). (d) Change of the number of electrons in the $\pi$-band as a function of pump-probe delay together with an exponential fit yielding a lifetime of $1.5\pm0.2$\,ps.}
  \label{figS8}
\end{figure}

\noindent In order to determine the transient chemical potential and electronic temperature of the graphene $\pi$-band we proceed as follows. First we integrate the photocurrent over the area between the two dashed lines in Fig. \ref{figS7}a. Then we fit the resulting EDC with a Fermi-Dirac distribution in the vicinity of the equilibrium chemical potential for all pump-probe delays. These fits describe the transient electronic distribution well at all times, indicating a rapid thermalization of the photo-excited carriers in the $\pi$-band in excellent agreement with literature \cite{BreusingPhysRevB2011, BridaNatCommun2013, GierzPhysRevLett2015, GierzJElectronSpectroscRelatPhenom2017}. The resulting shift of the chemical potential referenced with respect to the vacuum level $\mu_{e (vac)}$ and the transient electronic temperature T$_e$ are shown in Figs. \ref{figS8}a and b, respectively. The transient shift of the chemical potential referenced with respect to the graphene Dirac point $\mu_{e (ED)}$ (Fig. \ref{figS8}c) is then calculated by subtracting the band shift in Fig. 2b of the main text from the shift of $\mu_{e (vac)}$ in Fig. \ref{figS8}a. From Figs. \ref{figS8}b and c we can then directly calculate the change of the total number of electrons in the graphene layer (Fig. \ref{figS8}d) via

\begin{equation*}
\Delta n_e(t)=\int_{-\infty}^{\infty}{dE\,\rho(E)\left[f_{FD}(E,\mu(t),T(t))-f_{FD}(E,\mu_{0},T_{0})  \right]}
\end{equation*}

\noindent where $\rho(E)=\frac{2A_c}{\pi}\frac{|E|}{\hbar^2 v_F^2}$ is the density of states with the unit cell are $A_c=\frac{3\sqrt{3}a^2}{2}$ and the lattice constant $a=1.42$\,\AA. The equilibrium chemical potential $\mu_0=-0.3$\,eV was obtained from the high-resolution photoemission spectrum in Fig. 1a of the main text. The transient chemical potential is given by $\mu(t)=\mu_0+\Delta \mu_{e (ED)}(t)$. The equilibrium temperature is $T_0=T(t<0\,\text{ps})=300$\,K. The number of transferred holes shown in Fig. 4 of the main text is then given by $\Delta n_h(t)=-\Delta n_e(t)$.

\subsection*{Phase space}

\begin{figure}
	\center
		\includegraphics[width = 0.7\columnwidth]{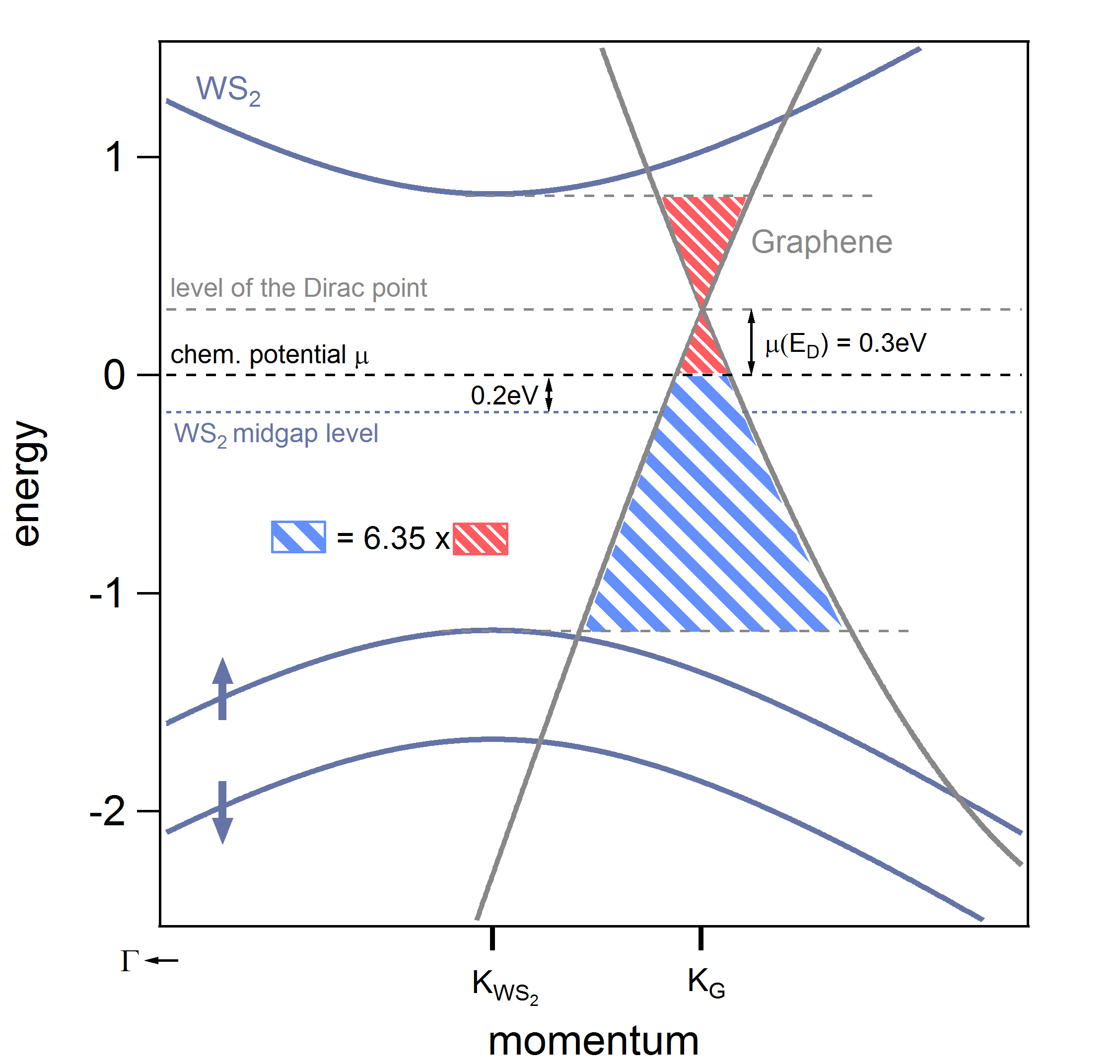}
  \caption{\textbf{Scattering phase space.} The WS$_2$ and graphene band structures are shown in blue and gray, respectively. The number of available electronic final states for electron and hole transfer are illustrated by red and blue areas, respectively.}
  \label{figS9}
\end{figure}

\noindent In Fig. \ref{figS9} we illustrate the number of available electronic final states for electron and hole transfer, respectively. The available number of final states (NOFS) for electrons and holes is given by

\begin{align*}
\text{NOFS}_e&=\int_{-\infty}^{E_{CB}}dE\,\rho(E)(1-f_{FD}(E,\mu,T))\\
\text{NOFS}_h&=\int_{E_{VB}}^{\infty}dE\,\rho(E)f_{FD}(E,\mu,T)
\end{align*}

\noindent where $E_{CB}$ ($E_{VB}$) is the position of the conduction band minimum (valence band maximum) of WS$_2$. The density of states $\rho(E)$ of graphene depends on the position of the Dirac point. The Fermi-Dirac distribution $f_{FD}(E,\mu,T)$ accounts for Pauli blocking. In the present case of the WS$_2$/graphene heterostructure the blue area (available electronic final states for hole transfer) is $\sim6$ times bigger than the red area (available electronic final states for electron transfer) which we expect to directly affect the corresponding charge transfer rates.

\noindent We would like to stress that a proper calculation of the scattering phase space requires detailed knowledge about the scattering mechanism (e.g. electron-electron or electron-phonon scattering) and should also include the phase space for the corresponding scattering partner.

\subsection*{Charge transfer exciton versus free electron-hole pairs}

\noindent In order to assess the stability of a putative interlayer exciton one needs to compare the exciton binding energy to the thermal energy of the carriers. Fermi-Dirac fits of the graphene $\pi$-band yield peak carrier temperatures of 1500\,K (see Fig. \ref{figS8}b). At this temperature only excitons with binding energies in excess of $k_BT=130$\,meV are expected to be stable. Neither the existence nor the binding energy of a putative interlayer exciton in WS$_2$/graphene is discussed in literature. We speculate that --- provided that the interlayer exciton exists --- its binding energy is most likely smaller than the binding energy of the A-exciton in monolayer WS$_2$ (which is 320\,meV \cite{ChernikovPhysRevLett2014}) due to enhanced screening by the metallic graphene layer.

\noindent Further, as shown in Fig. \ref{figS1} our samples consist of triangular WS$_2$ islands on top of the graphene layer with a total coverage of 40\% and a spacing between WS$_2$ islands of $1-10$\,$\mu$m. Due to the extreme surface sensitivity of ARPES we are only able to probe the graphene layer in the areas where it is uncovered. Also, the tr-ARPES measurements average over the area of the XUV spot with a diameter of $\sim300$\,$\mu$m. From the fact that the pump-probe signal of the graphene $\pi$-band of our WS$_2$/graphene heterostructure (main text) is clearly different from the one of pure graphene (Fig. \ref{figS5}) we conclude that all or a significant part of the holes that are transferred from WS$_2$ to graphene are rapidly delocalized over the complete graphene layer.

\noindent These points seem to be in favor of free electron-hole pairs rather than charge transfer excitons.


\end{document}